\documentclass{article}

\title{Bayesian inference and neural estimation of acoustic wave propagation}
\author{
    Yongchao Huang\textsuperscript{1}\thanks{yh522@cam.ac.uk, Department of Engineering, University of Cambridge} \and
    Yuhang He\textsuperscript{2}\thanks{yuhang.he@cs.ox.ac.uk, Department of Computer Science, University of Oxford} \and
    Hong Ge\textsuperscript{1}\thanks{hg344@cam.ac.uk, Department of Engineering, University of Cambridge}}
\date{May 2023}

\usepackage{biblatex}
\usepackage{graphicx}

\usepackage[title]{appendix}
\usepackage{etoolbox}
\BeforeBeginEnvironment{appendices}{\clearpage}

\usepackage[linesnumbered,ruled,vlined]{algorithm2e}
\usepackage{scrextend}
\usepackage{amsmath}
\usepackage{amssymb}

\usepackage{nicematrix}
\usepackage{float}
\usepackage{xcolor}
\usepackage{booktabs}
\usepackage{hyperref}

\PassOptionsToPackage{table}{xcolor}
\usepackage{xcolor}
\usepackage{subcaption}

\DeclareMathOperator*{\argmin}{arg\,min}

\begin{document}

\maketitle

\begin{abstract}
    In this work, we introduce a novel framework which combines physics and machine learning methods to analyse acoustic signals. Three methods are developed for this task: a Bayesian inference approach for inferring the spectral acoustics characteristics, a neural-physical model which equips a neural network with forward and backward physical losses, and the non-linear least squares approach which serves as benchmark. The inferred propagation coefficient leads to the room impulse response (RIR) quantity which can be used for relocalisation with uncertainty. The simplicity and efficiency of this framework is empirically validated on simulated data.    
\end{abstract}

\section{Introduction}

\paragraph{Bayesian inference} 
Bayesian inference seeks to infer the posterior distribution of some parameters of interest via the \textit{Bayes rule}:

\begin{equation} \label{Eq:posterior}
    p(\textbf{z}|\textbf{x}) = \frac{p(\textbf{x}|\textbf{z})p(\textbf{z})}{p(\textbf{x})}
\end{equation}

\noindent where $\textbf{z}=z_{1:m} \in \mathbb{R}^m$ are the latent variables that we are interested in performing inference on, $\textbf{x}=x_{1:n} \in \mathbb{R}^d$ are the observations, $p(\textbf{z})$ and $p(\textbf{z}|\textbf{x})$ are the prior and posterior distributions, respectively, $p(\textbf{x}|\textbf{z})$ is the data likelihood given $\textbf{z}$, and $p(\textbf{x})=\int_{\mathbb{R}^m} p(\textbf{x}|\textbf{z})p(\textbf{z})d\textbf{z}$ is the marginal likelihood, which can be hard to compute or intractable. Sampling methods such as Markov chain Monte Carlo (MCMC) \cite{MCMC_Andrieu} can be used to perform approximate inference \cite{VI_Blei}, and they provide  guarantees at stationary states. However, MCMC methods are known to be computational intensive \cite{MCMC_Robert} in large data settings \cite{MCMC_Bardenet}, and high dimensions \cite{MCMC_Barbos,MCMC_Yang}, or when the model is complex \cite{VI_Blei}. Bayesian inference has the advantage of sample efficiency and uncertainty quantification, it also offers the flexibility of continuous inference. Besides, it helps detect any inconsistency between the specified probabilistic model and data.

\paragraph{Sound wave propagation}

In time domain, the propagating waves are governed by the wave equation \footnote{This is, however, a very simplified version of more complex wave propagation theories which may involve wave inference, overlapping, reflection and refraction, etc.}

\begin{equation} \label{Eq:wave_equation_time_domain}
    \frac{\partial^2 p(x,t)}{\partial x^2} - \frac{1}{c^2} \frac{\partial^2 p(x,t)}{\partial t^2} = 0
\end{equation}

\noindent where $P(x,t)$ is the sound pressure which can be directly measured. Eq.\ref{Eq:wave_equation_time_domain} describes a wave varying in space and time (i.e. a 
 spatial-temporal model \footnote{Similar natural phenomena include heat transfer which is governed by the heat equation \cite{Heat_Widder}.}). We seek to find a solution $P(x,t)$ to this wave equation. In acoustics, we concern about the particle displacement u(x,t) which is normally proportional to the sound pressure $P(x,t)$. For simplicity, we assume the equality $P(x,t)=u(x,t)$. And the time domain solution $u(x,t)$ for a 1D propagating sound wave can be represented as \cite{Thesis_Yongchao}

\begin{equation} \label{Eq:wave_in_time_domain}
    u(x,t) = A e^{\pm \alpha x} e^{i(\omega t \pm \kappa x)}
\end{equation}

\noindent where $\omega$ is the angular frequency, $\alpha(\omega)$ is the frequency-dependent attenuation coefficient (also called damping coefficient), $\kappa$ is the wave number (also termed spatial velocity). Together they make the \textit{wave propagation coefficient} $\gamma(\omega)=\alpha(\omega)+i \kappa(\omega)$. Notes that \cite{Thesis_Yongchao}, $\alpha(\omega)$ is a positive even function, and $\kappa(\omega)$ is an odd function and positive for $\omega > 0$. The $\pm$ sign indicates the wave propagating in two directions. Energy dissipation is induced when wave propagating in, for example, viscoelastic medium, which is captured by the attenuation term $e^{\pm \alpha x}$, i.e. wave magnitude attenuation over the travel distance $x$; phase changes are determined by the term $e^{i(\omega t \pm \kappa x)}$. In acoustics, we also have the relation

\begin{equation} \label{Eq:wave_number}
    |\kappa(\omega)| = \frac{2\pi}{\lambda(\omega)} = \frac{\omega}{c(\omega)}
\end{equation}

\noindent where $c$ is the wave speed. Note that, $\alpha,\kappa,c$ are both frequency-dependent and material-dependent.  

More frequently, wave propagation phenomenon is analysed in frequency domain \cite{Thesis_Yongchao}. We could apply Fourier transform to the time domain wave equation, which gives \cite{Thesis_Yongchao}

\begin{equation} \label{Eq:wave_equation_frequency_domain}
    \frac{\partial^2}{\partial x^2} \Tilde{P}(x,\omega) + \frac{\omega^2}{c^2} \Tilde{P}(x,\omega) = 0
\end{equation}

\noindent and the frequency domain solution

\begin{equation} \label{Eq:wave_in_frequency_domain}
    \Tilde{P}(x,\omega) = \Tilde{P}_0(\omega) e^{-\gamma x} + \Tilde{P}'_0(\omega) e^{\gamma x}
\end{equation}

\noindent where $\Tilde{P}_0(\omega)$ and $\Tilde{P}'_0(\omega)$ are the magnitudes for each frequency components propagating in left and right directions (they are analogous to the initial time domain magnitude $A$ in Eq.\ref{Eq:wave_in_time_domain}). They can be determined by initial conditions.

Substituting the wave solution Eq.\ref{Eq:wave_in_frequency_domain} into the wave equation Eq.\ref{Eq:wave_equation_frequency_domain}, we have

\begin{equation} \label{Eq:wave_equation_frequency_domain_simplified}
    (\alpha^2 + i2\alpha\kappa)\Tilde{P}_0(\omega) e^{-(\alpha+i\kappa)x} + (\alpha^2 + i2\alpha\kappa)\Tilde{P}'_0(\omega) e^{(\alpha+i\kappa)x} = 0
\end{equation}

For the interest of our problem, we would like to just consider a one-direction propagating wave $\Tilde{P}(x,\omega)=\Tilde{P}_0(\omega) e^{-\gamma x}$, which simplifies Eq.\ref{Eq:wave_equation_frequency_domain_simplified} to be $(\alpha^2 + i2\alpha\kappa)\Tilde{P}_0(\omega) e^{-(\alpha+i\kappa)x}=0$, which is a complex-valued equality.

\paragraph{Notations} 

In this paper, we mainly concern about spatial and spectral notations. Frequency domain quantities are denoted with a tilde hat, e.g. $\Tilde{P}^m(x_{i1},\omega_{j})$. The superscript $m$ denotes measurement, distinguishing from its predicted counterpart without superscript. Speaker signal is always implied by the subscript "1", while receiver signal by "2". Whenever double indices is necessary, e.g. matrix entries, the first subscript indicates instance index, e.g. $x_{i1}$ corresponds to the position of speaker $i$. As each speaker-receiver pair gives a vector signal, aggregating all signals together forms a $N \times n$ matrix, where $N$ is the total number of signals and $n$ is the number of frequency components for example. Therefore, row $i$ accommodates the signal corresponding to the $(x_{i1},x_{i2})$ pair, and column $j$ refers to the magnitudes corresponding to (angular) frequency $\omega_j$. Vectors and matrices are represented in bold whenever convenient.

\section{Related Work}

\textbf{Room Acoustics Modelling} There are two main ways to model room acoustics: wave-based modelling~\cite{bilbao2017wave-based,pietrzyk1998computer,kleiner1993auralization-an,finitediff} and geometry-based modelling~(aka geometrical acoustics)~\cite{georoom_overview}. The wave-based modelling utilizes sound wave nature to model sound wave propagation, whereas geometry-based modelling treats sound propagation as optic rays. Typical geometry-based modelling methods include ray tracing~\cite{raytracing}, image source method (ISM) \cite{ISM}, beam tracing~\cite{beamtracing} and acoustic radiosity~\cite{exp_radiosity,inv_radiosity}. The main goal of room acoustics is to characterize the room reverberation effect, which consists of three main components: direct-path, specular-reflection and late-reverberation. 

\textbf{Neural Room Impulse Response~(RIR)} Some recent work~\cite{fast_rir,diff_acousgen,IR-GAN,TS-RIR,SDNet,NAF,richard2022deepimpulse,majumder2022fewshot} proposed to learn room impulse response~(RIR) with deep neural networks. However, they all assume massive source-to-receiver RIRs~(s2r-RIR) are available to train the model. Unlike these methods, SoundNeRirF learns implicit r2r-RIR from a set of 
robot recorded sounds at different positions, which are easier and more realistic to collect in real scenarios. 

\textbf{Relocalisation} Re-localising the agent in an environment through the received signals is of vital importance for various practical tasks, including positioning, mapping task in robotics~\cite{robot_reloc} and immersive virtual reality~(VR), online game experience~\cite{immersive_spatialaudio,3Dimmersive}. Existing methods extensively used data such as Laser scanner~\cite{robot_reloc,pcd_reloc}, RGB images~\cite{turkoglu2021visual} and spatial audio~\cite{SoundDoA,sounddet}.

\section{Estimation of acoustic characteristics}

It is the goal in many subjects (e.g. solid mechanics) to estimate the value of the frequency-dependent acoustics characteristics ($\gamma(\omega)$,$\alpha(\omega)$,$\kappa(\omega)$), as they are linked to the property (i.e. impedance, elasticity) of the medium the wave propagates in. Traditionally, wave propagation characteristics are estimated using least squares \cite{Thesis_Yongchao}. Here we introduce two further methodologies, namely Bayesian inference, and neural parameter estimation.

\subsection{Bayesian wave propagation analysis}
 
 The Bayesian approach not only allows encoding prior knowledge (e.g. a rough range of stiffness of the medium) into the estimation process, but also enables uncertainty quantification. The goal of performing Bayesian inference in frequency doamin is to obtain the posterior distributions of $\alpha(\omega)$ and $\kappa(\omega)$, using prior knowledge and data information. In Bayesian context, the problem can be formulated as: given a set of $\alpha(\omega)$ and $\kappa(\omega)$ values (e.g. those sampled from prior distributions), we can calculate the wave profile at $x_2$ as $\Tilde{P}(x_2,\omega)=\Tilde{P}^m (x_1,\omega) e^{-\gamma(\omega) (x_2-x_1)}$, and under certain noise variance $\sigma^2$, this predicted waveform should distribute around the measured wave profile, i.e. $\Tilde{P}(x_2,\omega) \sim \mathcal{N}(\Tilde{P}^m(x_2,\omega),\sigma^2)$, where the superscript $m$ denotes measurements. The same applies to $\Tilde{P}^m(x_1,\omega)$, if the wave were to travel inversely in space from $x_2$ to $x_1$. Putting them together, we have:

\begin{equation} \label{Eq:Bayesian_likelihood}
\begin{bmatrix}
\Tilde{P}^m(x_2,\omega) - \Tilde{P}^m(x_1,\omega) e^{-\gamma(\omega) (x_2-x_1)} \\
\Tilde{P}^m(x_1,\omega) - \Tilde{P}^m(x_2,\omega) e^{\gamma(\omega) (x_2-x_1)}
\end{bmatrix}
\sim
\mathcal{N}(
\begin{bmatrix}
0 \\ 0 
\end{bmatrix},
\begin{bmatrix}
\sigma^2 \\ \sigma^2
\end{bmatrix})
\end{equation}

\noindent which expands into a two-dimensional matrix if we flatten it, use each $(x_1,x_2)$ pair as row index and frequency $\omega$ as column index \footnote{So each row corresponds to an example instance, and each column corresponds to a particular frequency.}.

The wave equation serves as a constraint, it can be either loosely satisfied (i.e. used in the likelihood to allow noise perturbation) or used to filter out the samples as a hard constraint. Instead of inferring parameter values via sampling, alternatively an optimization person may thrive to minimize the LHS of Eq.\ref{Eq:Bayesian_likelihood}, in which nonlinear least squares \cite{OLS_Box} or maximum likelihood methods \cite{MLE_Eliason} can be applied.

\begin{algorithm}[H]
\caption{Bayesian inference of wave propagation coefficients (Metropolis-Hastings sampling as an example).}
\label{algo:BI_inference_of_wave_prop_coefficients}
\textbf{Inputs}: two time series waveform measurements $P^m(x_1,t_1)$, $P^m(x_2,t_2)$. \\

\vskip 0.06in
Calculate the travel distance $\Delta x=x_2-x_1$. Convert time domain measurements $P^m(x_1,t_1)$, $P^m(x_2,t_2)$ into frequency domain, obtain frequency domain data $\Tilde{P}^m(x_1,\omega)$, $\Tilde{P}^m(x_2,\omega)$. \\ 

Set prior distributions $p_r(\alpha)$, $p_r(\kappa)$ and noise level $p_r(\sigma)$. \\ 

Set sampling distributions $p_{lik}(\Tilde{P}(x_2,\omega)|\Tilde{P}^m(x_1,\omega),\alpha,\kappa)$ and $p_{lik}(\Tilde{P}(x_1,\omega)|\Tilde{P}^m(x_2,\omega),\alpha,\kappa)$ to be Gaussians with variance $\sigma^2$, where $\Tilde{P}(x_2,\omega) = \Tilde{P}^m(x_1,\omega) e^{-\gamma(\omega) (x_2-x_1)}$ and $\Tilde{P}(x_1,\omega) = \Tilde{P}^m(x_2,\omega) e^{\gamma(\omega) (x_2-x_1)}$ are the predicted wave profiles. \\

Choose proposal distributions $q_{\alpha}$, $q_{\kappa}$ and $q_{\sigma}$, e.g. symmetric Gaussians. \\

Sample an initial sample $\alpha_0$ and $\kappa_0$ from prior distributions. Note $\alpha_0$ and $\kappa_0$ are vectors with length equaling the number of frequencies.

For each iteration $l=0,1,2,...,L-1$, repeat: \\
    \begin{addmargin}[1em]{0em}%
    (1) generate the next sample candidate $(\alpha_{l+1},\kappa_{l+1},\sigma_{l+1})$ by sampling from the proposal distributions $q_{\alpha_{l+1}|\alpha_{l}}$, $q_{\kappa_{l+1}|\kappa_{l}}$ and $q_{\sigma_{l+1}|\sigma_{l}}$, respectively. \\ 
    (2) calculate the predicted wave profile $\Tilde{P}(x_2,\omega)=\Tilde{P}^m(x_1,\omega) e^{-\gamma (x_2-x_1)}$ and $\Tilde{P}(x_1,\omega)=\Tilde{P}^m(x_2,\omega) e^{\gamma (x_2-x_1)}$, where $\gamma=\alpha+i\kappa$. \\
    (3) evaluate the acquisition value: \\ $r=\frac{p_r(\alpha_{l+1}) \times p_r(\kappa_{l+1}) \times p_r(\sigma_{l+1}) \times p_{lik}(\Tilde{P}(x_2,\omega)|\Tilde{P}^m(x_1,\omega),\alpha_{l+1},\kappa_{l+1},\sigma_{l+1}) \times p_{lik}(\Tilde{P}(x_1,\omega)|\Tilde{P}^m(x_2,\omega),\alpha_{l+1},\kappa_{l+1},\sigma_{l+1})}{p_r(\alpha_{l}) \times p_r(\kappa_{l}) \times p_r(\sigma_{l}) \times p_{lik}(\Tilde{P}^m(x_1,\omega)|\Tilde{P}(x_2,\omega),\alpha_{l},\kappa_{l},\sigma_{l}) \times p_{lik}(\Tilde{P}^m(x_2,\omega)|\Tilde{P}(x_1,\omega),\alpha_{l},\kappa_{l},\sigma_{l})}$. \\
    (4) calculate the left hand side of the wave equation: $LHS=(\alpha^2 + i2\alpha\kappa)\Tilde{P}(x_1,\omega) e^{-(\alpha+i\kappa) \Delta x}$. \\
    (5) generate a uniform random number $r' \in [0,1]$. If $r' \leq r$, $\alpha > 0$ and $LHS=0$, accept the candidate. Else, reject the candidate and set $\alpha_{l+1}=\alpha_l$, $\kappa_{l+1}=\kappa_l$ and $\sigma_{l+1}=\sigma_l$ instead.
    \end{addmargin}

\textbf{\textit{Return}} sample trajectories $\alpha_0, \alpha_1,...\alpha_{L-1}$ and $\kappa_0, \kappa_1,...\kappa_{L-1}$. \\
\end{algorithm}

The detailed procedure of Bayesian inference is described in Algorithm.\ref{algo:BI_inference_of_wave_prop_coefficients}, in which we have used the conditions $LHS=0$ and $\alpha>0$ as constraints to select sample values. As mentioned before, instead of using $LHS=0$ as a filtering criteria in Step 7.5, we could pack the LHS values into the likelihood (i.e. the sampling distribution), which turns the condition into a loose constraint. For example, if we choose Gaussian distribution as the sampling distribution, we have $(\Tilde{P}_t(x_2,\omega), LHS) \sim \mathcal{N}(\textbf{0}, (\sigma, \sigma') \otimes \textbf{I})$ \footnote{$\otimes$ here denotes Kronecker product.}, where $\sigma'$ is the noise tolerance for LHS.

\subsection{Neural parameter estimation}

\begin{figure}[ht]
    \centering
    \includegraphics[width=0.7\columnwidth]{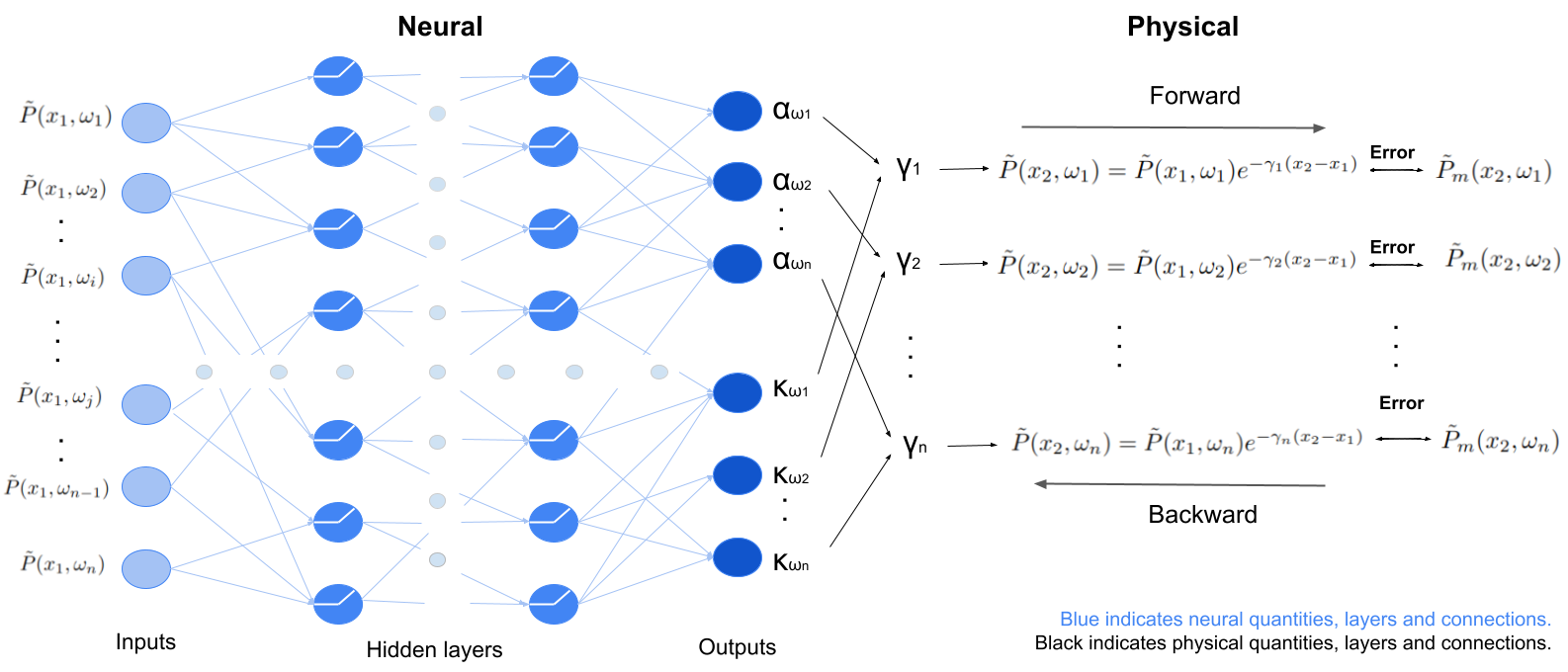}
    \caption{The neural-physical model architecture.}
    \label{fig:neural_net_diagram}
\end{figure}

\paragraph{General principle} Observing the non-linear nature of wave propagation, instead of using Bayesian inference for parameter estimation, we could also build a neural network for predicting the parameter values. These neural estimated physical quantities, i.e. the attenuation coefficient $\alpha$ and phase speed $\kappa$, are then plugged into the physics, i.e. the assumed solution $\Tilde{P}(x_2,\omega)=\Tilde{P}(x_1,\omega) e^{-\gamma (x_2-x_1)}$ to the spectral wave equation Eq.\ref{Eq:wave_equation_frequency_domain}, to yield a predicted waveform profile at the receiver position. The gap between the predicted waveform, typically the sum of magnitude discrepancies over the frequency spectrum across training instances, can be minimized by adjusting the network weights via back-propagation. The optimal set of network weights would output the wave propagation coefficient such that, given the input wave (i.e. the speaker waveform), the predicted output (i.e. the receiver waveform) would match the ground truth measurement. Fig.\ref{fig:neural_net_diagram} sketches the design of the the neural-physical model (i.e. the neural-physical model).

\paragraph{Loss design} 

A wise design of the loss functional could accelerate the learning phase and improve accuracy. As the physical process is reversible in space, that is, if the received signal is sent back to the speaker, then same input wave should be derived using the neural net predicted propagation coefficient. By inverting the frequency domain solution $\Tilde{P}(x,\omega)=\Tilde{P}_0(\omega) e^{-\gamma x}$, we can obtain the inverse process dynamics $\Tilde{P}(x_1,\omega) = \Tilde{P}(x_2,\omega) e^{\gamma (x_2-x_1)}$, which has also been used in the Bayesian likelihood. This motivates the forward-backward loss design, where the forward loss accounts fro the mismatch between the received signals, while the backward loss measures the discrepancies between the input signals if we reverse the physical process. We can further implement the principle of minimum entropy of the predicted coefficients across all instance predictions, as the environment or medium in which the wave propagating in stays static over the experimental time horizon. With these principles in mind, the following target functional is constructed to guide the training of the neural-physical model:

\begin{equation} \label{Eq:neaural_physical_model_loss_func}
\begin{split}
\text{Loss}(\theta;x_1,x_2) = \frac{1}{N \times n} \sum_{i=1}^N  \sum_{j=1}^n \{ [\Tilde{P}^m_i(x_2,\omega_j)-\Tilde{P}_i(x_2,\omega_j,\theta)]^2 \\
+ [\Tilde{P}^m_i(x_1,\omega_j)-\Tilde{P}_i(x_1,\omega_j,\theta)]^2 + [C - \textit{I} (\frac{1}{N} \sum_{i=1}^{N} C_{i,\cdot})^T ]^2 \}
\end{split}
\end{equation}

\noindent where $\theta$s are the neural net weights. $\omega$ is the angular frequency, $n$ is the total number of symmetric frequency components after discrete Fourier transform (DFT). $N$ is the number of training data point (each data point is a row vector of Fourier magnitudes with length $n$). $\Tilde{P}^m(x_1)$ and $\Tilde{P}^m(x_2)$, both with size $(N,n)$, are the Fourier magnitude matrix of the \textit{measured} signals, $\Tilde{P}(x_1)$ and $\Tilde{P}(x_2)$ are their corresponding \textit{predicted} counterparts \footnote{Note that, as all inputs and outputs are in frequency domain, these matrix entries are mostly complex-valued. For two complex numbers, when evaluating their discrepancy, contributions from both real and imaginary parts are aggregated.}. $I$ is a $N$-length column vector with identity elements. $C_{N \times n}$ is the propagation coefficient matrix from the outputs of the neural network:

\begin{equation}
    C_{N \times n} = 
    \begin{bmatrix}
        \alpha_{1,1} & \alpha_{1,2} &\dots & \alpha_{1,n} & \kappa_{1,1} & \kappa_{1,2} &\dots & \kappa_{1,n} \\
        \alpha_{2,1} & \alpha_{2,2} &\dots & \alpha_{2,n} & \kappa_{2,1} & \kappa_{2,2}  &\dots &\kappa_{2,n} \\
        \vdots  &\vdots  &\vdots &\vdots  &\vdots  &\vdots  &\vdots  &\vdots \\
       \alpha_{N,1} & \alpha_{N,2} &\dots & \alpha_{N,n} & \kappa_{N,1} & \kappa_{N,2}  &\dots &\kappa_{N,n} \\
    \end{bmatrix}
\end{equation}

\noindent with each column corresponds to a frequency and each row represents the coefficients prediction for input instance $i$. 

The first two terms in the curl brackets of Eq.\ref{Eq:neaural_physical_model_loss_func} measure the forward and backward discrepancies, respectively. The third term measures the row concentration: each row of the propagation coefficient coefficient matrix is de-meaned by the row average to form a residual matrix; minimizing the size of the residual matrix encourages all rows to be close to each other. Squared Frobenius norm (i.e. the $L_2$ distance) is used to measure the sizes of all three residual matrices (note all matrix squares in Eq.\ref{Eq:neaural_physical_model_loss_func} denote element-wise squares).

\paragraph{Uncertainty quantification} In general, a neural network only gives point estimate; we can however using stochastic gradient descent (SGD) to train multiple neural networks, where each network is exposed to partial data, and obtain slightly different predictions for all input instances, which yields credible prediction intervals. This emsembling approach can help the neural network generalise (potentially increase bias as well) but could be computationally intensive. In the scenario of small or medium size data set, we demonstrate that it's efficient and robust in later experiments. From a Bayesian perspective, the emsembling method can be regarded as applying uniform prior to data batches. In an extreme case where all selected data batches are exclusive and a consistent weight initialisation strategy, e.g. $p(\theta_k) \sim \mathcal{N}(0,1)$, is applied, we can approximately obtain the posterior distribution of the weights as: $p(\theta|D) \propto \sum_k p(D_k|\theta_k) p(\theta_k)$ where $k$ is a batch index and $\theta = \bigcup_{k} \theta_k$.

\subsection{Non-linear least squares} To minimize the LHS in Eq.\ref{Eq:Bayesian_likelihood} (same as the first two terms in Eq.\ref{Eq:neaural_physical_model_loss_func}), we use non-linear least squares without regularization. We can first apply the log trick for exponential functions, which gives $\log \Tilde{P}^m(x_{i2},\omega) - \log \Tilde{P}^m(x_{i1},\omega) \approx -\gamma(\omega) (x_{i2}-x_{i1})$. Expanding the equation to host rows as $(x_{i1},x_{i2})$ pair instances and use frequency as column index, we obtain the matrix representation $\Delta \log \Tilde{P} \approx - \Delta \textbf{x} \gamma^T$, where $\Delta \log \Tilde{P}_{N \times n} (x_{i1},x_{i2},\omega_j) = \Tilde{P}_{N \times n}(x_{i2},\omega_j) - \Tilde{P}_{N \times n}(x_{i1},\omega_j)$ is the measurements difference matrix, $\Delta \textbf{x}_{N \times 1}=\{x_{i2}-x_{i1}\}_{i=1}^N$ is the collection of travel distances. $\gamma_{N \times 1}(\omega_i)=\alpha(\omega_i)+i \kappa(\omega_i)$ is a vector of propagation coefficients corresponding to each of the $n$ frequency components. 

As $\Delta \log \Tilde{P}_{N \times n}$ is matrix-valued, we first observe that Eq.\ref{Eq:log_least_squares} is a system of linear equations: $\Delta \log \Tilde{P}_i = - \Delta x_i \times \gamma^T$. If we make $x$ as the diagonal entries of a diagonal matrix $\Delta X_{N \times N}$, and construct a new matrix $\Gamma_{N \times n}$ by dulicating $\gamma^T$ along each row, then the following matrix least squares problem can be formulated:

\begin{equation} 
\label{Eq:log_least_squares}
    \hat{\gamma} = \argmin_{\gamma} \lVert \Delta \log \Tilde{P}_{N \times n} - (-\Delta X_{N \times N} \Gamma_{N \times n}) \rVert_2^2
\end{equation}

\noindent which can be found by solving the normal equations, yielding 
\[\hat{\Gamma}=-(\Delta X^T \Delta X)^{-1} \Delta X^T \Delta \log \Tilde{P}=-(\Delta X)^{-1} \Delta \log \Tilde{P}
\]
This least squares solution implies that $\hat{\Gamma}_{i,\cdot}=-\frac{\Delta \log \Tilde{P}_{i,\cdot}}{x_{i2}-x_{i1}}, i=1,2,...N$, which, not surprisingly, fits exactly the proposed wave equation solution $\Tilde{P}^m(x_{i2},\cdot)=\Tilde{P}^m(x_{i1},\cdot) e^{-\gamma (x_{i2} - x_{i1})}$. In an ideal scenario (i.e. noise free), all rows of $\hat{\Gamma}$ should be equal; when noise is present, for each case of the input pairs, we obtain an estimated version of $\gamma(\omega)$, and least squares in this way yields an interval estimate. This uncertainty comes from the inconsistency in the numerator and denominator of the least squares solution, which may be induced by noise; this is different from the Bayesian credible interval which is generated directly by sampling and rejection (noise is used in the likelihood and the MH step though).

\section{Estimating room impulse response (RIR)}

The room impulse response $RIR(x,t)$ is a convolutional basis signal in time domain which, after convolved with the input signal $P(x,t)$, can be used to generate the output time domain signal:

\begin{equation} \label{Eq:RIR_time_domain}
    P(x_1,t) \circledast RIR(x_2-x_1, t) = P(x_2,t)
\end{equation}

\noindent where $\circledast$ here denotes convolution operator. $RIR(x,t)$ has the same size as the input signal (e.g. fixed sampling rate), and is generally an indicator of the acoustics characteristics of the room (e.g. air density), given the ambient conditions (e.g. temperature, humidity, etc). Eq.\ref{Eq:RIR_time_domain} is another representation of the wave propagation phenomenon: given the room geometry and conditions, it assumes a fixed convolutional operand $RIR(x,t)$ between two fixed locations. It reflects the state of the medium in which the wave propagates in.

Applying Fourier transform to Eq.\ref{Eq:RIR_time_domain} gives the frequency domain representation\footnote{$\odot$ here denotes element-wise multiplication (also known as the Hadamard product or Schur product). Here the time-frequency transform trick applies: convolution in time domain is equivalent to multiplication in frequency domain, and vice versa.}:

\begin{equation} \label{Eq:RIR_frequency_domain}
    \Tilde{P}(x_1,\omega) \odot \Tilde{RIR}(x_2-x_1,\omega) = \Tilde{P}(x_2,\omega)
\end{equation}

\noindent from which we obtain $\Tilde{RIR}(x_2-x_1,\omega) = \Tilde{P}(x_2,\omega) ./ \Tilde{P}(x_1,\omega)=e^{-\gamma (x_2-x_1)}$, where $./$ denotes element-wise division (i.e. the Hadamard division). We observe that, if $\Tilde{P}(x_1,\omega) := \textbf{1}$, which corresponds to a Dirac delta impulse in time domain \footnote{The frequency domain response of the Dirac delta function is a constant with a magnitude of 1 at all frequencies.}, then we have $ \Tilde{RIR}(x_2-x_1,\omega) = \Tilde{P}(x_2,\omega)$. That is, $\Tilde{RIR}(x_2-x_1,\omega)$ equals the response (output of the system) measured at $x_2$, when the excitation (input of the system) at $x_1$ is a Dirac unit impulse. This implies that, with he learned parameters ($\alpha$,$\kappa$) of the system\footnote{System here, in analogy to control terminologies, is used to represent the medium between $x_1$ and $x_2$; more precisely, it is the wave propagation path.}, we can purposely inject a Dirac unit impulse into the system and obtain $\Tilde{RIR}(x_2-x_1,\omega)$ by analysing the received signal at $x_2$. The time domain $RIR(x_2-x_1,t)$ can be obtained by inverting the frequency domain signal.

If we have measurements $\Tilde{P}(x_1,\omega)$ and $\Tilde{P}(x_2,\omega)$ at two arbitrary locations $x_1$ and $x_2$, we can obtain 

\begin{equation} \label{Eq:frequency_RIR_estimate}
    \Tilde{RIR}(x_2-x_1,\omega) = \Tilde{P}(x_2,\omega) ./ \Tilde{P}(x_1,\omega)=e^{-\gamma (x_2-x_1)}
\end{equation}

\noindent Note that, in a homogeneous environment, i.e. the properties of the medium (e.g. density) in which wave propagates are constant over time and space (e.g. humidity is not changing over time or temperature), the wave propagation coefficient $\gamma$ remains the same for all locations (i.e. orientation isotropic). The RIR, however, is a function of both the wave propagation coefficient and the wave travel distance $x_2-x_1$ (the absolute coordinates are not relevant).

\section{Relocalisation}

Here we demonstrate an application of the aforementioned method for relocalisation in robotics. We have the scenario where a moving robot would like to evaluate its distance from known speaker positions via collecting and processing information from these fixed-position speakers. Given two wave profiles $\Tilde{P}_1(\omega)$ and $\Tilde{P}_2(\omega)$, we can use Eq.\ref{Eq:RIR_frequency_domain} to inversely calculate the distance $\delta x = x_2 - x_1$:

\begin{equation} \label{Eq:dist_estimate}
    \Delta x = x_2 - x_1 = -\frac{1}{\gamma} \ln \Tilde{RIR}(x_2-x_1,\omega) = -\frac{1}{\gamma} \ln \frac{\Tilde{P}(x_2,\omega)}{\Tilde{P}(x_1,\omega)} 
\end{equation}

\noindent For 1D navigation, knowing the speaker location $x_1$ (e.g.  a fixed speaker), we are able to estimate the current position $x_2$ of the mobile robot equipped with a receiver. Due to noise, however, the resulted $\Delta x$ ratio may not be constant. We can therefore obtain a distribution of $\Delta x$ and take a mean estimate \footnote{Alternatively, we can take the mode as an estimate.}:

\begin{equation} \label{Eq:Delta_x_mean}
    \widehat{\Delta x} = \frac{1}{n} \sum_{i=1}^n -\frac{1}{\gamma} \ln \frac{\Tilde{P}(x_2,\omega_i)}{\Tilde{P}(x_1,\omega_i)}
\end{equation}

\noindent where $n$ is the number of frequency components in the Fourier spectrum.

\begin{figure}[ht]
    \centering
    \includegraphics[width=.65\linewidth]{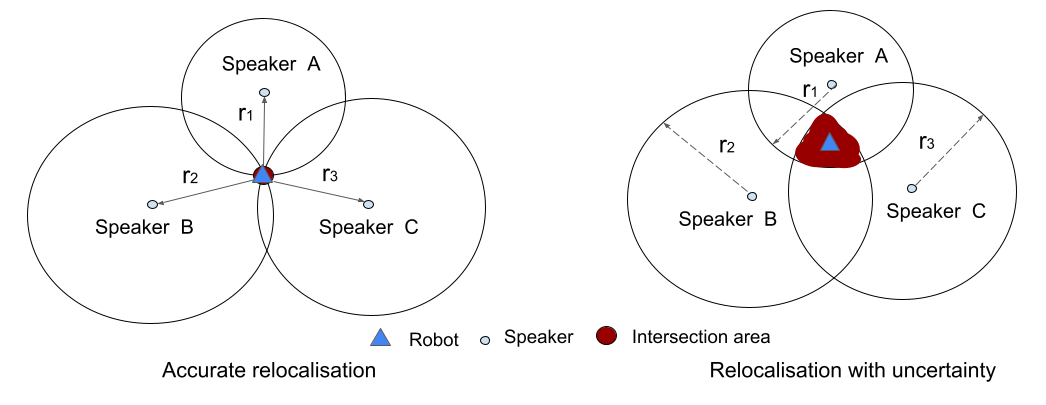}
    \caption{2D/3D relocolisation diagram.}
    \label{fig:2D_relocalisation}
\end{figure}

For 2D navigation problem \footnote{In 3D, the circles become balls, but the same principle follows.}, however, we can only locate $x_2$ on the circle with radius $\Delta x$ centering the speaker position; to locate the absolute location of the robot, we need signals from extra two speakers, calculate each radius and take the intersection as the search area, see Figure.\ref{fig:2D_relocalisation}. 

\section{Experiments} \label{sec:experiments}

\begin{figure}[ht]
    \centering
    \begin{subfigure}{0.40\textwidth}
        \centering
        \includegraphics[width=\linewidth]{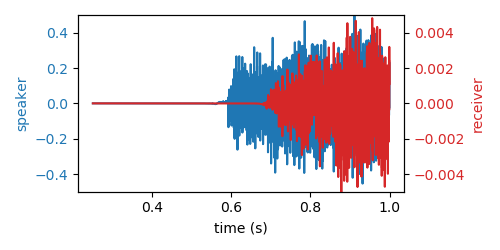}
        \caption{Time domain.}
        \label{fig:speaker_receiver_time_domain_signals}
    \end{subfigure}
    \begin{subfigure}{0.40\textwidth}
        \centering
        \includegraphics[width=\linewidth]{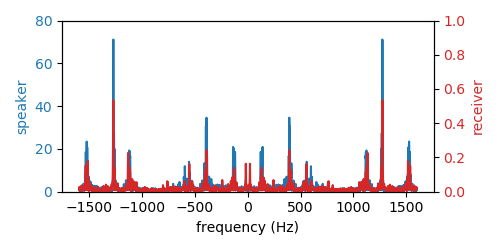}
        \caption{Frequency domain.}
        \label{fig:speaker_receiver_frequency_domain_signals}
    \end{subfigure}
    \caption{The Bayesian training speaker-receiver signal pair. Speaker coordinate: (-9.308,0.021,-0.270), receiver coordinate: (-1.994,0.072,-1.439), $\Delta x = 7.408$.}
    \label{fig:speaker_receiver_signals}
\end{figure}

\paragraph{Experimental setup} 
We use a room simulator \textit{SoundSpaces 2.0}~\cite{chen22soundspaces2} to simulate a homogeneous room with \textit{Matterport3D} dataset~\cite{Matterport3D}. The simulated room maximally resembles a real room environment. In the simulated environment, a static speaker with fixed position constantly sends out a fixed signal, and 9 receivers, located at different positions, collect their signals without inference. All signals are 1 second long with sampling rate 16k. For Bayesian inference, only one pair of of speaker-receiver signals, shown in Fig.~\ref{fig:speaker_receiver_signals}, is used as training data; for the neural-physical model, 8 signal pairs are used in training, while the test set consists of the 1 remaining signal pair. Least squares utilizes just 1 pair of signals for estimation.

\subsection{Bayesian inference results}

The probabilistic model is specified as follows: multivariate Gaussian priors are used for $\mathbf{\alpha}$ and $\mathbf{\kappa}$, and half normal distribution is applied for noise variance, i.e. 
\[
    \mathbf{\alpha} \sim \mathcal{N}(\mathbf{1}, \mathbf{\textit{I}}), 
    \mathbf{\kappa} \sim \mathcal{N}(\mathbf{0}, 10^2\mathbf{\textit{I}}),
    \sigma \sim \text{\textit{HalfNormal}}(std=1)
\]

\begin{figure}[ht]
  \centering
  \begin{subfigure}[b]{0.3\textwidth}
    \centering
    \includegraphics[width=\textwidth]{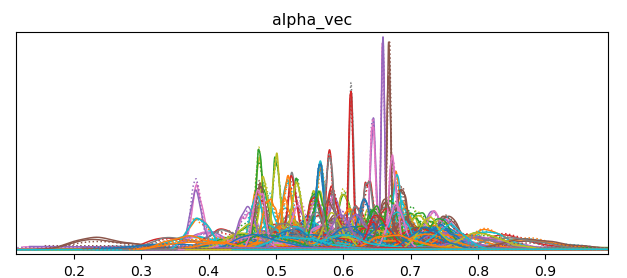}
    \caption{$\alpha$}
    \label{fig:MCMC_sampling_1}
  \end{subfigure}
  \hfill
  \begin{subfigure}[b]{0.3\textwidth}
    \centering
    \includegraphics[width=\textwidth]{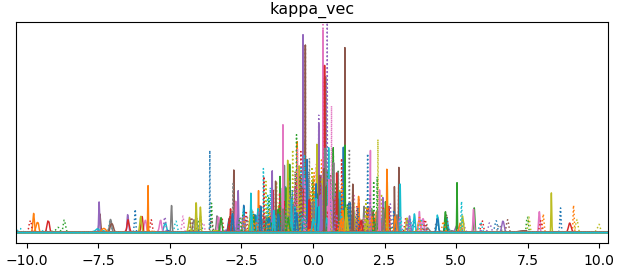}
    \caption{$\kappa$}
    \label{fig:MCMC_sampling_2}
  \end{subfigure}
  \hfill
  \begin{subfigure}[b]{0.3\textwidth}
    \centering
    \includegraphics[width=\textwidth]{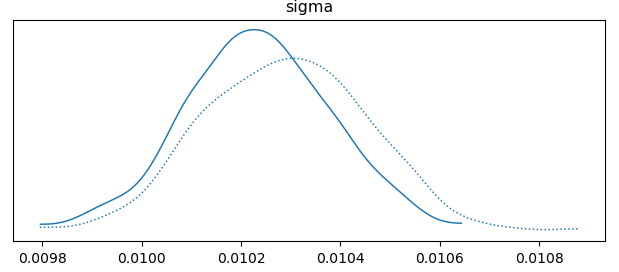}
    \caption{$\sigma$}
    \label{fig:MCMC_sampling_3}
  \end{subfigure}
  \caption{MCMC sample trajectories. Two MCMC chains, each with 1000 iterations, are generated.}
  \label{fig:MCMC_sampling}
\end{figure}

We use the forward likelihood specified previously in the first half of Eq.\ref{Eq:Bayesian_likelihood}. The \textit{NUTS} sampler \cite{NUTS_Hoffman}, a variant Hamiltonian Monte Carlo method, is used to perform Bayesian inference on the posterior density of $(\alpha,\kappa,\sigma)$. Table.\ref{Tab:Bayesian_inference_hybrid_mode} shows the statistics of 6 parameters \footnote{As the number of frequency components is large, we arbitrarily select these 6 parameters.} and the noise variance. The convergence diagnostic $\hat{R}$ (i.e. the Gelman-Rubin statistic) \cite{MCMC_Vehtari}, which evaluates the between and within-chain mixing, concentrates around 1, and the effective sample size (ESS, the number of "independent" samples) is sizeable. In Fig.\ref{fig:MCMC_sampling}, the damping coefficient $\alpha$ is positive, and the wave number symmetrically distributed around 0, which is consistent with the physics. This suggests that MCMC inference of the posterior has been successful. 

\begin{table}[ht]
\begin{center}
\small
\begin{tabular}{ m{2.0cm} m{1.1cm} m{1.1cm} m{1.1cm} m{1.1cm} m{1.1cm} m{1.1cm} m{1.1cm}} 
  \toprule 
  Quantity &$\alpha_{0}$ &$\alpha_{500}$ &$\alpha_{1000}$ &$\kappa_{0}$ &$\kappa_{500}$ &$\kappa_{1000}$ &$\sigma$ \\ 
  \midrule
  Mean  &1.561           &1.603            &1.536         &0.348          &-0.689            &0.156    &0.010 \\ 
  Std  &0.704           &0.709            &0.728        &10.789           &9.818           &10.133    &0.000 \\   
  $\hat{R}$ &1.000           &1.000            &1.000         &1.010           &1.000            &1.010    &1.050 \\  
  ESS (tail)  &569         &493          &321       &569         &750          &486  &255 \\  
  \bottomrule
\end{tabular}
\caption{Summary statistics of MCMC samples.}
\label{Tab:Bayesian_inference_hybrid_mode}
\end{center}
\end{table}

\begin{figure}[ht]
    \centering
    \includegraphics[width=0.55\linewidth]{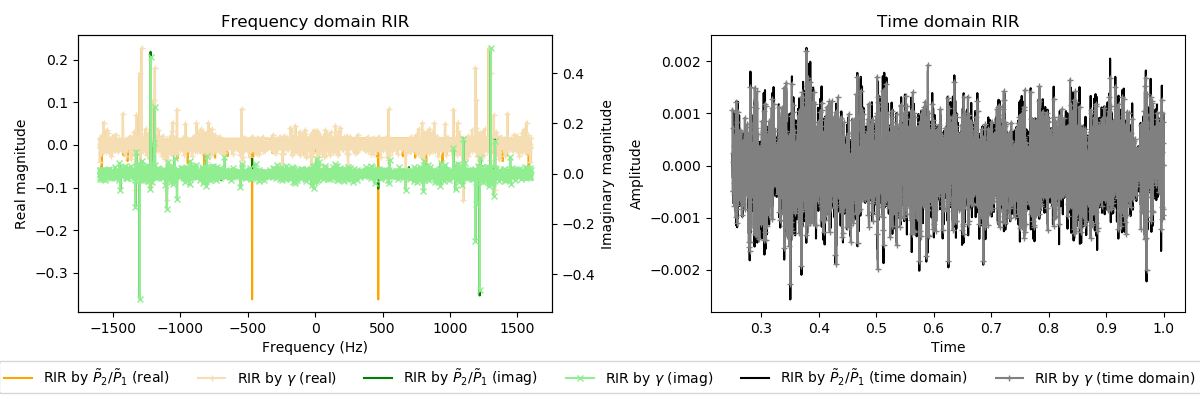}
    \caption{Posterior MAP estimated RIR for the training signal pair.}
    \label{fig:Bayesian_inference_RIR_training}
\end{figure}

\begin{figure}[ht]
    \centering
    \begin{subfigure}{0.55\textwidth}
        \centering
        \includegraphics[width=\linewidth]{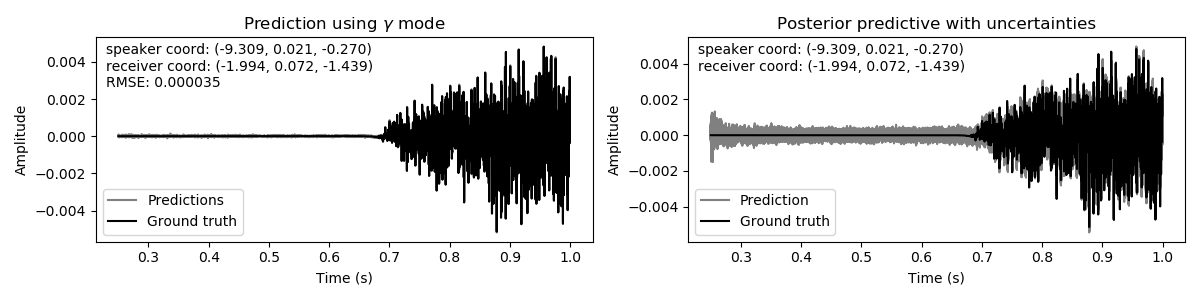}
        \caption{Predicting the training wave profile.}
        \label{fig:predict_training_wave}
    \end{subfigure}
    \\
    \begin{subfigure}{0.55\textwidth}
        \centering
        \includegraphics[width=\linewidth]{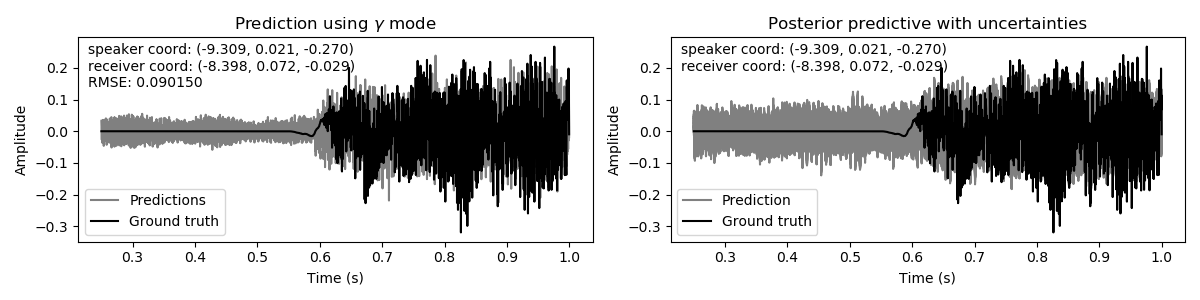}
        \caption{Predicting a test wave profile.}
        \label{fig:predict_test_wave}
    \end{subfigure}
    \caption{Predictions using posterior MAP and samples from MCMC inference results.}
    \label{fig:Bayesian_inference_time_domain_predictions}
\end{figure}

After obtaining the propagation coefficient $\gamma(\omega)=\alpha(\omega)+i\kappa(\omega)$ using sample modes, we plug it into Eq.\ref{Eq:frequency_RIR_estimate} and obtain RIR, as shown in Fig.\ref{fig:Bayesian_inference_RIR_training}. To validate Bayesian learning, given the speaker signal $\Tilde{P}(x_1,\omega)$, we use the wave solution $\Tilde{P}(x_2,\omega)=\Tilde{P}(x_1,\omega) e^{-\gamma(\omega) (x_2-x_1)}$ to predict the wave profile $\Tilde{P}(x_2,\omega)$ and compare it with ground truth. This is shown in Fig.\ref{fig:Bayesian_inference_time_domain_predictions}, where we observe a good fit of the training wave, while the predicted test wave is reasonably good. The uncertainties, generated by sampling $\gamma$ from the posterior, give credible interval for the predictions.

\begin{figure}[ht]
    \centering
    \begin{subfigure}{0.25\textwidth}
        \centering
        \includegraphics[width=\linewidth]{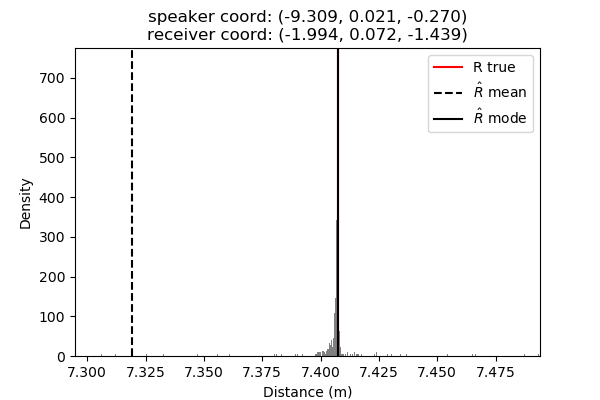}
        \caption{Test receiver 1.}
        \label{fig:Bayesian_predict_radius_R10}
    \end{subfigure}
    \begin{subfigure}{0.25\textwidth}
        \centering
        \includegraphics[width=\linewidth]{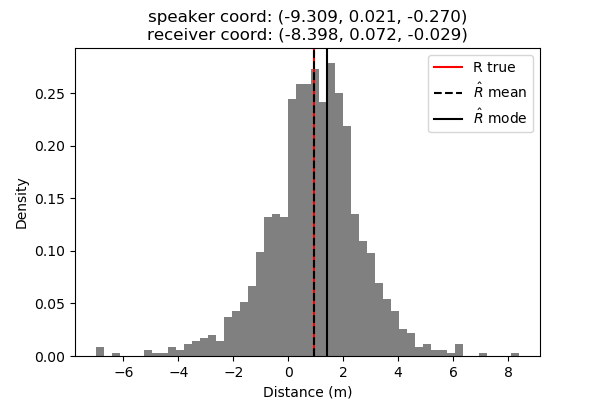}
        \caption{Test receiver 2.}
        \label{fig:Bayesian_predict_radius_R1}
    \end{subfigure}
    \caption{Bayesian inference: estimating the distance between speaker and receiver using MAP of $\gamma$.}
    \label{fig:Bayesian_inference_radius_prediction}
\end{figure}

Having the two waves measured at the speaker and the robot positions, we can use Eq.\ref{Eq:dist_estimate} to obtain a distribution of the distance $\Delta x$; uncertainty is also propagated from parameter estimation. Fig.\ref{fig:Bayesian_inference_radius_prediction} shows the estimated distances for two receivers. In both cases, with noise presence, we still observe good consistency with ground truth. In parallel, we present the results using posterior mean estimates for $\gamma$ in Appendix.\ref{app:Bayesian_inference_further_results}.

\subsection{Neural parameter estimation results} 

\begin{figure}[ht]
    \centering
    \includegraphics[width=0.8\linewidth]{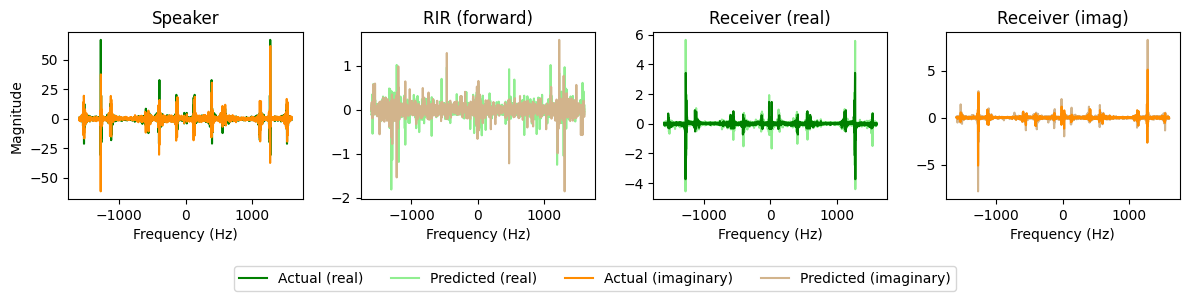}
    \caption{Neural-physical model performance in frequency domain on the training sample with coordinate (-3.987,0.072,-1.271).}
    \label{fig:NN_traing_performance_frequency_domain}
\end{figure}

\begin{figure}[ht]
    \centering
    \includegraphics[width=0.55\linewidth]{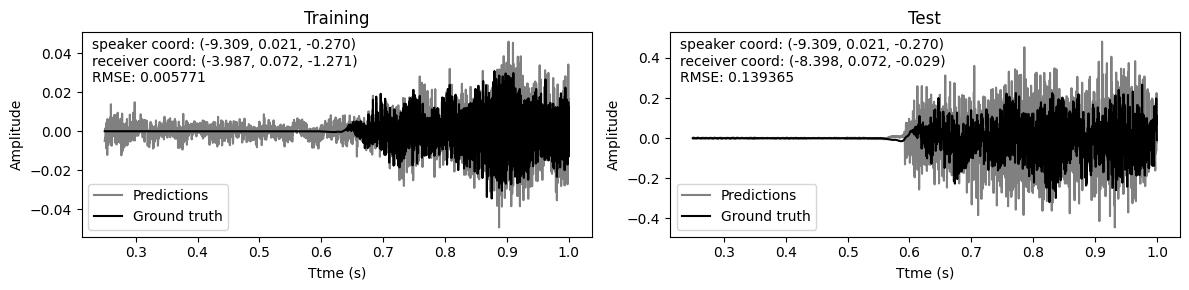}
    \caption{Neural-physical model performances in time domain on training and test samples.}
    \label{fig:NN_training_performance_time_domain}
\end{figure}

\begin{figure}[ht]
    \centering
    \begin{subfigure}{0.25\textwidth}
        \centering
        \includegraphics[width=\linewidth]{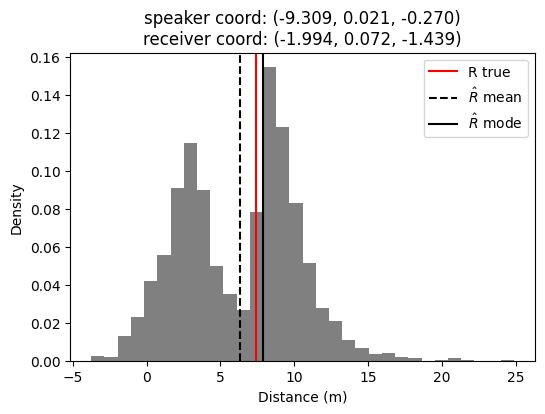}
        \caption{ Test receiver 1.}
        \label{fig:NN_predict_radius_R10}
    \end{subfigure}
    \begin{subfigure}{0.25\textwidth}
        \centering
        \includegraphics[width=\linewidth]{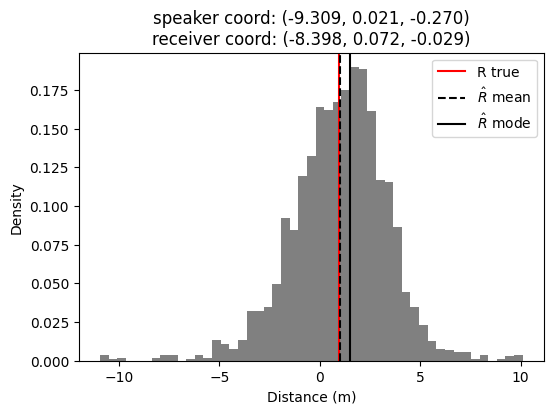}
        \caption{Test receiver 2.}
        \label{fig:NN_predict_radius_R1}
    \end{subfigure}
    \caption{Estimating the distance between speaker and receiver using neural estimated $\gamma$ (posterior MAP).}
    \label{fig:NN_radius_prediction}
\end{figure}

\begin{figure}[ht]
    \centering
    \begin{subfigure}{0.25\textwidth}
        \centering
        \includegraphics[width=\linewidth]{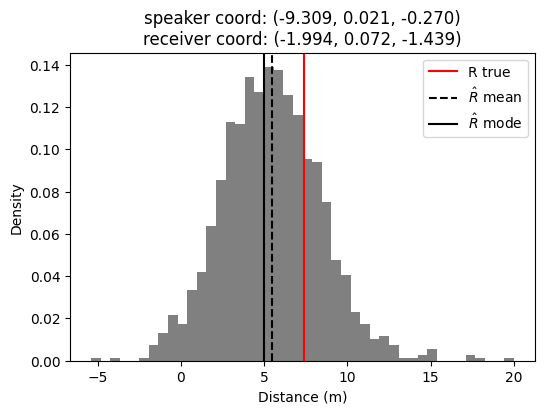}
        \caption{ Test receiver 1.}
        \label{fig:least_squares_predict_radius_R10}
    \end{subfigure}
    \begin{subfigure}{0.25\textwidth}
        \centering
        \includegraphics[width=\linewidth]{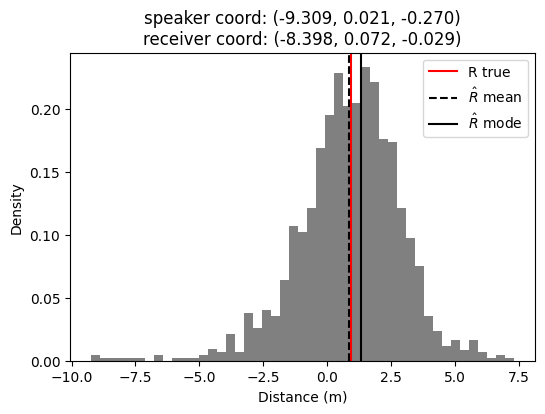}
        \caption{Test receiver 2.}
        \label{fig:least_squares_predict_radius_R1}
    \end{subfigure}
    \caption{Estimating the distance between speaker and receiver using least squares estimated $\gamma$.}
    \label{fig:least_squares_radius_prediction}
\end{figure}

When training the neural-physical model, we have available the speaker and receiver positions, as well as their wave profiles. A fully connected, multi-layer perceptron (MLP) network, with \textit{ReLU} activations and layer sizes (4800, 128, 256, 256, 128, 4800), is constructed to consume the speaker spectral waveform and outputs the propagation coefficient. Eight wave profiles are used as training set and one remains as the test instance. The loss converges after 500 epochs with a learning rate of 1e-4. Fig.\ref{fig:NN_traing_performance_frequency_domain} shows the model performance, over one of the training sample, in frequency domain; we observe reasonably good match between the actual and predicted spectral signals. Fig.\ref{fig:NN_training_performance_time_domain} further compares the training and test performances in time domain. We observe that the test performance, quantified by the RMSE metric, is slightly worse than the Bayesian method presented in Fig.\ref{fig:predict_test_wave}. Using the neural estimated wave propagation coefficient, we are able to derive the distance between any two speaker-receiver pair. A comparison of the distance estimates are made in Fig.\ref{fig:NN_radius_prediction} (neural estimation), Fig.\ref{fig:least_squares_radius_prediction} (least squares) and Table.\ref{Tab:comparison_distance_methods} in which we also include results from Bayesian mean estimation (Appendix.\ref{app:Bayesian_inference_further_results}) and neural autoencoder estimation (Appendix.\ref{app:autoencoder}). Among these three distance estimation methods, Bayesian inference wins in both cases, although the other two methods also show comparable performances.


\begin{figure}[ht]
    \centering
    \includegraphics[width=0.70\linewidth]{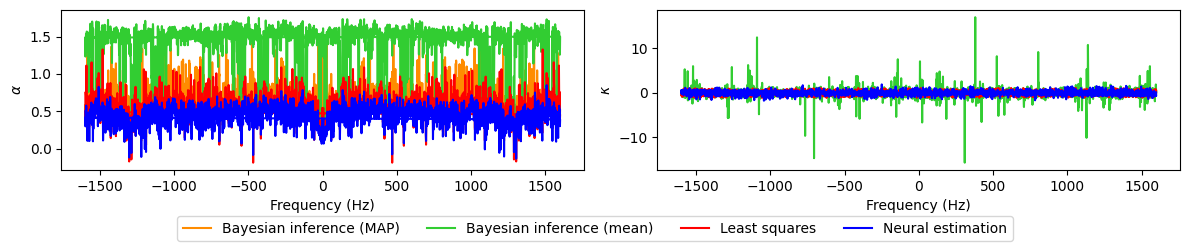}
    \caption{Comparison of the estimated propagation coefficient using three methods. Fully connected MLP architecture is used for neural estimation. Least squares estimations are derived using the receiver signal at coordinate (-3.987,0.072,-1.271).}
    \label{fig:NN_compare_gamma}
\end{figure}

Finally, we compare the estimated propagation coefficient in Fig.\ref{fig:NN_compare_gamma}, also presented are the least squares estimations obtained using a single pair of speaker-receiver signals. It is observed that, neural estimation and least squares strictly preserve the symmetries in $\alpha$ and $\kappa$ ($\alpha$ is symmetric \textit{w.r.t} y-axis, while $\kappa$ is symmetric about the origin.); Bayesian inferred coefficients, as observed in Fig.\ref{fig:MCMC_sampling}, loosely satisfy the symmetry. The neural network learns the symmetry from the inputs as well as by the loss constraint. Also, neural estimation gives estimations on a similar scale as that of least squares, while Bayesian inference gives different values depending on whether mode or mean sample values are chosen. The Bayesian MAP values coincide with neural estimation and least squares, while mean posterior values deviate from others. These frequency-dependent values, i.e. $0 \leq \alpha <5$ and $|\kappa|<100$, are sensible for wave propagation in air; they are dependent on some intrinsic properties of the propagating medium such as temperature, humidity, impedance and resistance.

\begin{table}[ht]
\centering
\footnotesize
\begin{tabular}{ m{3.5cm} m{1cm} m{1cm} m{1cm} m{1cm} m{1cm} m{1cm}}
  \toprule 
  & \multicolumn{3}{c}{Test receiver 1} & \multicolumn{3}{c}{Test receiver 2} \\ 
  \cmidrule(lr){2-4} \cmidrule(lr){5-7}
  Statistics & Mean & Mode & Std & Mean & Mode & Std \\ 
  \midrule
  Bayesian inference (MAP) &\textbf{7.319} &\textbf{7.407} &\textbf{0.466} &\textbf{0.936} &1.425 &1.715 \\ 
  Bayesian inference (mean) &2.749 &2.896 &1.311 &0.416 &0.487 &0.855 \\ 
  Neural estimation (MLP) &6.290 &7.883 &3.955 &0.986 &1.509 &2.395 \\ 
  Neural estimation (Autoencoder) &6.308 &7.623 &3.880 &0.961 &1.869 &2.375 \\ 
  Least squares &5.458 &4.998 &2.968 &0.854 &\textbf{1.318} &2.016 \\
  Ground truth &\textcolor{red}{\textbf{7.407}} & \centering - & \centering - &\textcolor{red}{\textbf{0.943}} & \centering - & - \\
  \bottomrule
\end{tabular}
\caption{Comparison of the three distance estimation methods.}
\label{Tab:comparison_distance_methods}
\end{table}

\paragraph{Discussion}

The Bayesian method provides a principled framework for learning and inference. It learns from small amount of data and achieve impressive performance in time domain prediction (Fig.\ref{fig:Bayesian_inference_time_domain_predictions}) and distance estimation (Fig.\ref{fig:Bayesian_inference_radius_prediction}), this is attributed to the embedding of prior knowledge (i.e. priors representing the physical constraints) and the contribution from data-evidenced likelihood. Beyond priors and sample efficiency, it also offers uncertainty quantification, i.e. parameter uncertainty propagates to prediction time (e.g. Fig.\ref{fig:Bayesian_inference_time_domain_predictions}), which is advantageous particularly in safety-critical scenarios (e.g. fire alarming, self-driving) where the posterior can supply better risk assessment over all possible outcomes compatible with observations and thus more informed decisions \cite{Bayesian_Taka}. The neural-physical method considers the forward and backward physical loss by design (see Fig.\ref{fig:neural_net_diagram} and Eq.\ref{Eq:neaural_physical_model_loss_func}), which might be of advantage compared to black-box deep learning approaches. Learning with physical constraints is efficient as it makes most use of data and may avoid trajectories which otherwise be a waste; uncertainty can be yielded by emsembling multiple networks which are trained on subsets of data. We have also applied weights as hyper-parameters to the loss components; weighting the loss components changes the landscape of the loss function. With the log transform trick, least squares for the wave propagation problem enjoys an intrinsic solution compatible with physical intuition, it's fast and usually gives point estimation (the uncertainty in this case comes naturally from physics). All three methods give wave propagation estimates, as well as the resulting distance estimation, with different degrees of accuracy (Fig.\ref{fig:NN_compare_gamma}); the inference quality is assessed by the match between the recovered signal and measurements, as well as their physical meanings.

\paragraph{Limitations and future work} More work can be done in the future to improve the existing approaches. First, the environment is assumed to be homogeneous, and wave solution in this work doesn't involve wave inference (e.g. overlapping); we can increase the complexity of the wave propagation equation to account for multi-direction wave propagation, reflection and absorption, with the expectation of achieving better accuracy. Second, scalable Bayesian methods are demanded for large data setting. One can aggregate the training data for Bayesian inference, expecting better generalized capacity and robustness; the same can be achieved with Bayesian online inference when data is collected in a streaming manner. Continuous Bayesian model updates is computationally advantageous compared to least squares which requires retraining on the entire dataset. Further, Bayesian inference results can tell when the probabilistic model is mis-specified, i.e. when the data is inconsistent with the model assumptions. Therefore, apart from model calibration, we can use existing model to detect data heterogeneity and environment changes. For deep spectral learning, accuracy may be improved by introducing complex-valued neural network which utilizes complex derivative (e.g. Wirtinger calculus \cite{complexNN_torch}) to correlate the real and imaginary parts. One can also weight the training instances with prior beliefs about data quality and noise levels to improve modelling. Using symmetric architectures (e.g. auto-encoder) may be advantageous in some specific problem setting (e.g. temporally and spatially reversible processes). Besides, future work can explore sequential learning methods such as deep learning based sequence-to-sequence models \cite{Kuznetsov2018FoundationsOS} for benchmarking.

\section{Conclusion}

In this work, we combine machine learning and physics to perform inference on acoustic characteristics in frequency domain. Three parameter and distance estimation methods, i.e. Bayesian inference, neural-physical estimation, and non-linear least squares, are demonstrated to be efficient and effective in this task. The Bayesian method enjoys learning from small data, encoding prior knowledge and quantifying uncertainty; the neural network and least squares learning is fast and yields desired symmetry for the physical quantities of concern. All three approaches give satisfying learning outcome. The physics-informed machine learning models achieve sample efficiency by making the most use of available data (e.g. the forward and backward losses), has improved expressivity (e.g. quality of the learning outcome can be assessed by the physical intuition) and interpretability (e.g. Bayesian prior, physical meanings of learning quantities), it also accelerates the learning process, e.g. by embedding the physical loss in neural network training. This work provides a simple and efficient framework for applications such as robot relocalisation with acoustic-only measurements.

\section*{Acknowledgements}
Yongchao would like to acknowledge the support and knowledge transfer from his PhD years with Prof. Clive Siviour. The authors also thank Xianda Sun for insightful discussion on training the neural network, as well as valuable support from CBL Cambridge.

\clearpage
\printbibliography

\clearpage
\appendix

\section{Bayesian inference: results using posterior mean estimates} \label{app:Bayesian_inference_further_results}

\begin{figure}[ht]
    \centering
    \includegraphics[width=0.55\linewidth]{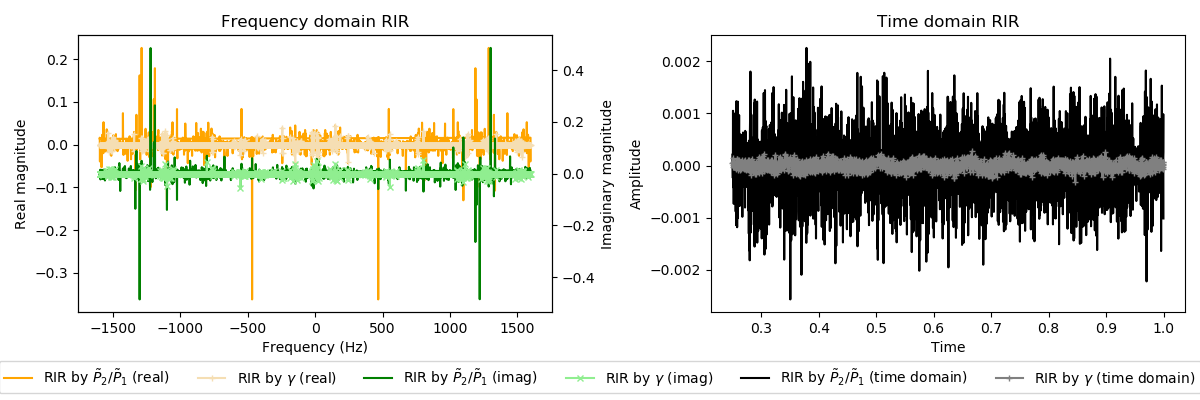}
    \caption{Mean posterior estimated RIR for the training signal pair.}
    \label{fig:Bayesian_inference_RIR_training_meanPosterior}
\end{figure}

\begin{figure}[ht]
    \centering
    \begin{subfigure}{0.55\textwidth}
        \centering
        \includegraphics[width=\linewidth]{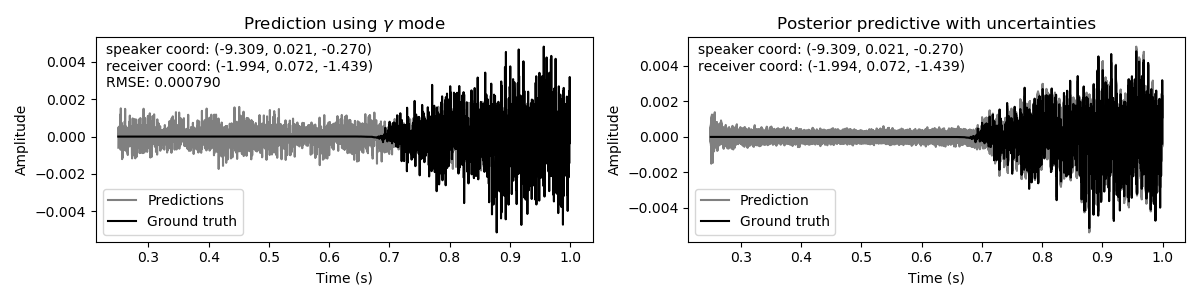}
        \caption{Predicting the training wave profile.}
        \label{fig:predict_training_wave_meanPosterior}
    \end{subfigure}
    \\
    \begin{subfigure}{0.55\textwidth}
        \centering
        \includegraphics[width=\linewidth]{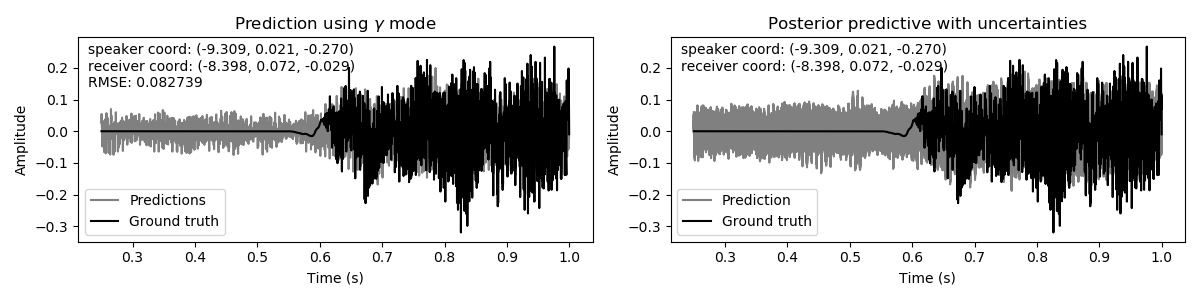}
        \caption{Predicting a test wave profile.}
        \label{fig:predict_test_wave_meanPosterior}
    \end{subfigure}
    \caption{Predictions using posterior mean and samples from MCMC inference results.}
    \label{fig:Bayesian_inference_time_domain_predictions_meanPosterior}
\end{figure}

\begin{figure}[ht]
    \centering
    \begin{subfigure}{0.25\textwidth}
        \centering
        \includegraphics[width=\linewidth]{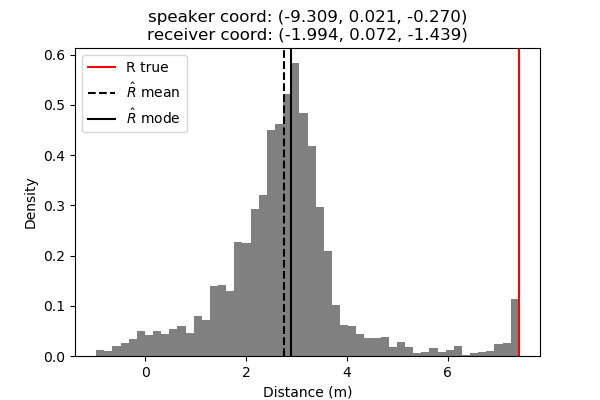}
        \caption{Test receiver 1.}
        \label{fig:Bayesian_predict_radius_R10_meanPosterior}
    \end{subfigure}
    \begin{subfigure}{0.25\textwidth}
        \centering
        \includegraphics[width=\linewidth]{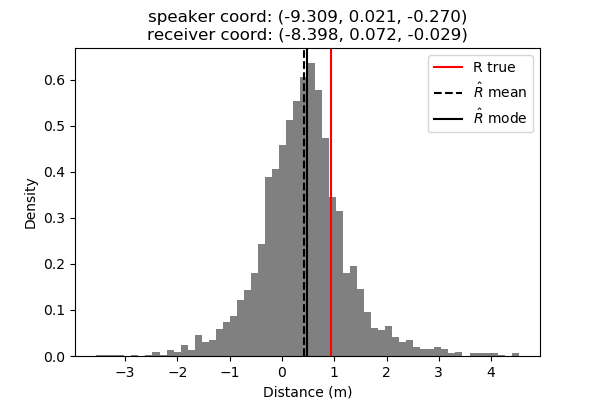}
        \caption{Test receiver 2.}
        \label{fig:Bayesian_predict_radius_R1_meanPosterior}
    \end{subfigure}
    \caption{Bayesian inference: estimating the distance between speaker and receiver using mean of $\gamma$.}
    \label{fig:Bayesian_inference_radius_prediction_meanPosterior}
\end{figure}

Probabilistic models are implemented in \textit{PyMC} \cite{PyMC_Salvatier}; there are other probabilistic programming packages available, e.g. \textit{Stan} \cite{Stan_Carpenter} and \textit{Turing.jl} \cite{Turing_Hong}. The NUTS sampler \cite{NUTS_Hoffman}, which is a variant of Hamiltonian Monte Carlo (HMC) sampler \cite{HMC_Duane}, is used to perform inference on $(\alpha,\kappa,\sigma)$. We sample 2 chains for each parameter, with chain length 1000; the initial 600 warm-up samples are discarded. Sampling is performed using 2 processors, which typically takes less than 5 minutes (sampling time is largely affected by the volume of data used, e.g. sampling rate, single signal or aggregated signals). Here we present the predictions of RIR (Fig.\ref{fig:Bayesian_inference_RIR_training_meanPosterior}), time domain waveforms (Fig.\ref{fig:Bayesian_inference_time_domain_predictions_meanPosterior}) and distance (Fig.\ref{fig:Bayesian_inference_radius_prediction_meanPosterior}) using mean posterior estimates from forward mode inference.

\section{Neural parameter estimation: results using auto-encoder architecture} \label{app:autoencoder}

\begin{figure}[ht]
    \centering
    \includegraphics[width=0.40\textwidth]{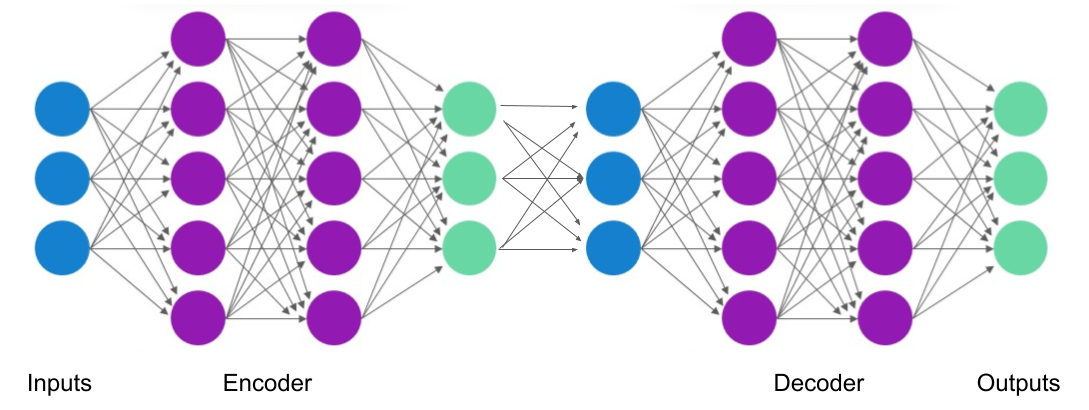}
    \caption{An illustrative diagram of the autoencoder architecture.}
    \label{fig:autoencoder_architecture}
\end{figure}

In parallel to the MLP architecture, here we employ the autoencoder architecture for neural parameter estimation. The autoencoder sequentially consists of an encoder and a decoder with symmetric, fully connected layers of sizes (4800, 128, 256, 256, 128, 128, 256, 256, 128, 4800) and \textit{ReLU} activations. An conceptual diagram of the autoencoder architecture is shown in Fig.\ref{fig:autoencoder_architecture}. The input is the speaker spectral waveform and output the propagation coefficient. Eight wave profiles are used as training data and one remains as the test instance. \textit{RMSProp} \cite{Ruder2016AnOO}, a variant of the gradient descent method, is used to update the network weights. In total 500 epochs with a learning rate of 1e-4 are used. 

\begin{figure}[ht]
    \centering
    \begin{subfigure}{0.25\textwidth}
        \centering
        \includegraphics[width=\linewidth]{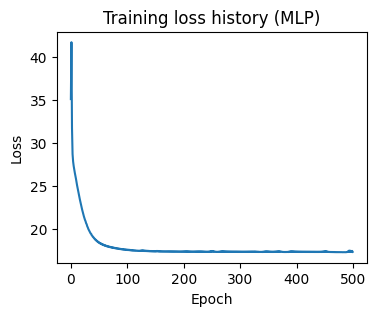}
        \caption{MLP.}
        \label{fig:MLP_training_loss_curve}
    \end{subfigure}
    \begin{subfigure}{0.25\textwidth}
        \centering
        \includegraphics[width=\linewidth]{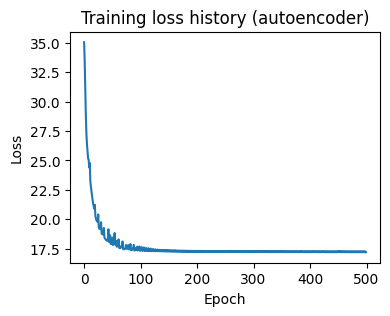}
        \caption{Autoencoder.}
        \label{fig:autoencoder_training_loss_curve}
    \end{subfigure}
    \caption{Neural network training loss history.}
    \label{fig:NN_training_loss_curves}
\end{figure}

\begin{figure}[ht]
    \centering
    \includegraphics[width=0.8\linewidth]{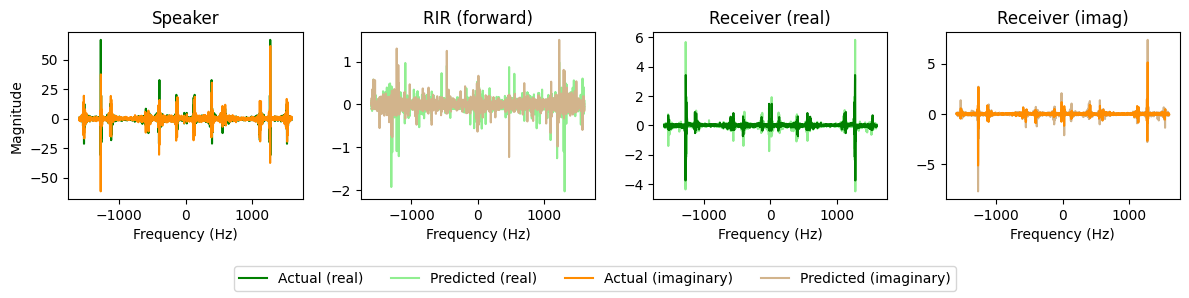}
    \caption{Neural-physical model performance in frequency domain on the training sample with coordinate (-3.987,0.072,-1.271).}
    \label{fig:autoencoder_traing_performance_frequency_domain}
\end{figure}

\begin{figure}[ht]
    \centering
    \includegraphics[width=0.55\linewidth]{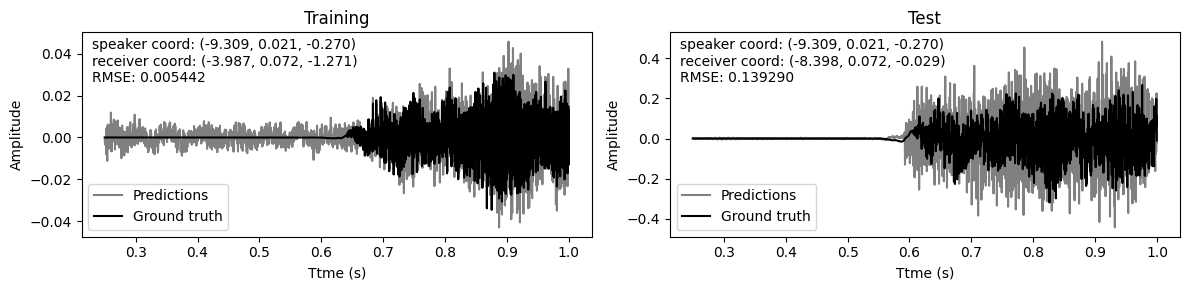}
    \caption{Neural-physical model performances in time domain on training and test samples.}
    \label{fig:autoencoder_training_performance_time_domain}
\end{figure}

\begin{figure}[H]
  \centering
  \begin{subfigure}{0.22\textwidth}
    \centering
    \includegraphics[width=\linewidth]{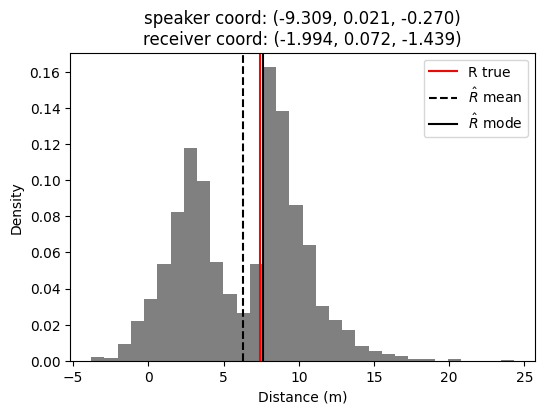}
    \caption{Test receiver 1.}
    \label{fig:autoencoder_predict_radius_R10}
  \end{subfigure}
  \begin{subfigure}{0.22\textwidth}
    \centering
    \includegraphics[width=\linewidth]{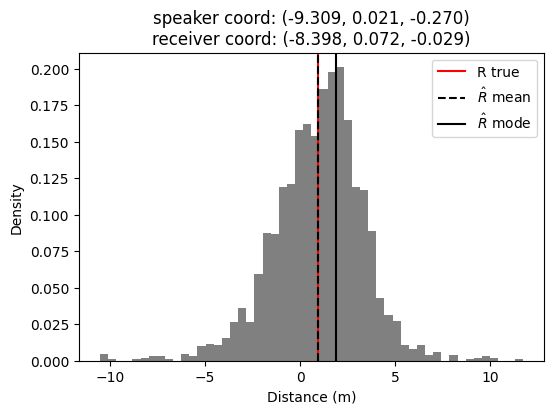}
    \caption{Test receiver 2.}
    \label{fig:autoencoder_predict_radius_R1}
  \end{subfigure}
  \caption{Estimating the distance between the speaker and receiver using neural (autoencoder) estimation of $\gamma$.}
  \label{fig:autoencoder_radius_prediction}
\end{figure}

\begin{figure}[H]
    \centering
    \includegraphics[width=0.70\linewidth]{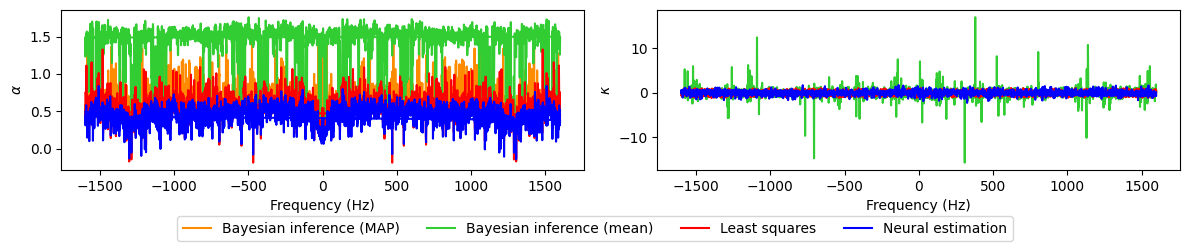}
    \caption{Comparison of the estimated propagation coefficient using three methods. Autoencoder architecture is used for neural estimation. Least squares estimations are derived using the receiver signal at coordinate (-3.987,0.072,-1.271).}
    \label{fig:autoencoder_compare_gamma}
\end{figure}

Fig.\ref{fig:NN_training_loss_curves} shows the training history, it is observed that the empirical losses stabilize after 200 epochs for both MLP and autoencoder scenarios. Fig.\ref{fig:autoencoder_traing_performance_frequency_domain} and Fig.\ref{fig:autoencoder_training_performance_time_domain} confirm the match between actual and predicted signals. Similar to MLP, the test RMSE of autoencoder is observed to be slightly worse than the Bayesian method. The estimated distance, obtained using the mean posterior $\gamma$, is shown in Fig.\ref{fig:autoencoder_radius_prediction}. A comparison of the estimated wave propagation coefficient is shown in Fig.\ref{fig:autoencoder_compare_gamma}, and the comparison of the estimated distances in Table.\ref{Tab:comparison_distance_methods}. It is seen that, the Bayesian MAP estimates give cloest results to the ground truth.

\section{Software packages and code availability}

All experiments are implemented in \textit{Python}. Fourier transforms are performed using \textit{SciPy} \cite{scipy_Jones}, Bayesian inference is implemented in \textit{PyMC3} \cite{PyMC_Salvatier}, while neural networks are constructed with \textit{PyTorch} \cite{NEURIPS2019_9015}. Random effects are minimized by setting seeds for reproducibility. Main packages are listed in Table.~\ref{tab:package_versions}. All software are available open source. Codes will be available at: $\text{https://github.com/YongchaoHuang/wave}$.

\begin{table}[h]
\centering
\begin{tabular}{|p{2.5cm}|p{2.5cm}|}
\hline
\textbf{Package} & \textbf{Version} \\ \hline
PyTorch           & v2.0.1                  \\ \hline
Numpy            & v1.22.4                 \\ \hline
Pandas           & v1.5.3                 \\ \hline
PyMC3            & v3.11.2                 \\ \hline
SciPy            & v1.10.1                 \\ \hline
\end{tabular}
\caption{Version of main packages (conda environment)}
\label{tab:package_versions}
\end{table}

\end{document}